# A THEORY OF DARK ENERGY THAT MATCHES DARK MATTER

Huai-Yu Wang[a]

Department of Physics, Tsinghua University, Beijing 100084, China

**Abstract:** In this paper, a theory of dark energy is proposed that matches dark matter. The relativistic quantum mechanics equations reveal that free particles can have negative energies. We think that the negative energy is the dark energy which behaviors as dark photons with negative energies. In this work, the photon number states are extended to the cases where the photon number can be negative integers, called negative integer photon states, the physical meaning of which are that the photons in such a state are of negative energy, i.e., dark photons. The dark photons constitute dark radiation, also called negative radiation. The formulism of the statistical mechanics and thermodynamics of the dark radiation is presented. This version of dark energy is of negative temperature and negative pressure, the latter regarded as responsible for the accelerate expansion of the universe. It is believed that there is a symmetry of energy-dark energy in the universe. In our previous work, the theory of the motion of the matters with negative kinetic energy was presented. In our opinion, the negative kinetic energy matter is dark matter. In the present work, we demonstrate that the dark substances absorb and release dark energy. In this view, the dark matter and dark energy match. Therefore, there is a symmetry of matter-energy match and dark matter-dark energy match in the universe. We present the reasons why the negative kinetic energy systems and negative radiation are dark to us.

[a] wanghuaiyu@mail.tsinghua.edu.cn

**Résumé**: Dans cet article, une théorie de l'énergie noire est proposée qui correspond à la matière noire. Les équations relativistes de la mécanique quantique révèlent que les particules libres peuvent avoir des énergies négatives. Nous pensons que l'énergie négative est l'énergie noire qui se comporte comme des photons noirs avec des énergies négatives. Dans ce travail, les états du nombre de photons sont étendus aux cas où le nombre de photons peut être des nombres entiers négatifs, appelés états de photons entiers négatifs, dont la signification physique est que les photons dans un tel état sont d'énergie négative, c'est-à-dire des photons sombres. Les photons noirs constituent le rayonnement noir, également appelé rayonnement négatif. Le formalisme de la mécanique statistique et de la thermodynamique du rayonnement noir est présenté. Cette version de l'énergie noire est de température négative et de pression négative, cette dernière étant considérée comme responsable de l'accélération de l'expansion de l'univers. On pense qu'il existe une symétrie énergie-énergie sombre dans l'univers. Dans nos travaux précédents, la théorie du mouvement des matières à énergie cinétique négative a été présentée. À notre avis, la matière à énergie cinétique négative est de la matière noire. Dans le présent travail, nous démontrons que les substances sombres absorbent et libèrent de l'énergie noire. Dans cette vue, la matière noire et l'énergie noire correspondent. Par conséquent, il existe une symétrie de correspondance matière-énergie et de correspondance matière noire-énergie noire dans l'univers. Nous présentons les raisons pour lesquelles les systèmes d'énergie cinétique négative et le rayonnement négatif nous sont obscurs.

## I. INTRODUCTION

In our previous work,[1-7] the fundamental equations of quantum mechanics (QM), including nonrelativistic and relativistic ones, were carefully inspected. The relativistic quantum mechanics equations (EQMEs), Dirac equation and Klein-Gordon equation, have both positive kinetic energy (PKE) and negative kinetic energy (NKE) branches, which is a remarkable feature. Let us briefly retrospect the concept of the NKE.

In classical mechanics, the expression of kinetic energy in Newtonian mechanics is

$$K_{(+)} = \frac{1}{2m} \boldsymbol{p}^2. \tag{1}$$

In special relativity, a free particle has energy

$$E_{(+)} = \sqrt{m^2 c^4 + c^2 \boldsymbol{p}^2}. \tag{2}$$

The subscript $(+)$ in (1) means PKE and that in (2) means that this energy is positive. Equation (2) is valid for any momentum value. When the momentum value is very low, (2) can be approximated to be (1) plus a static energy $mc^2$.

In QM, the kinetic energy in the Schrödinger equation is positive. Indeed, according to the Schrödinger equation, the kinetic energy is of the form of (1). In a region where the potential $V$ is less than the particle's energy $E$, $E > V$, the kinetic energy is positive. People naturally extended the application of the Schrödinger equation to regions where $E < V$. It was pointed out[1] that whether this extension was correct or not had never been experimentally quantitatively verified nor theoretically derived.

After the Schrödinger equation, the RQMEs were proposed. From the RQMEs, the energy of a free particle can be

$$E_{(\pm)} = \pm\sqrt{m^2 c^4 + c^2 \boldsymbol{p}^2}. \tag{3}$$

That is to say, there are two energy branches. The positive one is the same as that in classical mechanics, (2). The negative branch was new:

$$E_{(-)} = -\sqrt{m^2 c^4 + c^2 \boldsymbol{p}^2}. \tag{4}$$

Equation (4) is valid for any momentum value. When the momentum value is very low, (4) can be approximated to be

$$K_{(-)} = -\frac{1}{2m} \boldsymbol{p}^2 \tag{5}$$

plus a constant $-mc^2$. Equation (5) is just opposite in sign to (1), and is the NKE. The two branches in Eq. (3) are called PKE and NKE branches, respectively.

It is well known that the Schrödinger equation could be derived from the RQMEs by taking low-momentum approximation. The procedure is as follows. Denote the wave function of the Dirac equation by $\Psi$. Let

$$\Psi = \psi_{(+)} e^{-imc^2 t/\hbar} \tag{6}$$

be substituted into the Dirac equation and take low-momentum approximation. Then, one obtains that the wave function $\psi_{(+)}$ obeys the Schrödinger equation. The transformation (6) actually made use of the PKE branch $E_{(+)}$ and the resulted Schrödinger equation for the function $\psi_{(+)}$ applied to regions $E > V$, because the kinetic energy in the Schrödinger equation has always the form of (1).

It was put forth[1,4,5] that there could be another transformation,

$$\Psi = \psi_{(-)} e^{imc^2 t/\hbar}. \tag{7}$$

After (7) is substituted into the Dirac equation and low-momentum approximation is made, one achieves the NKE Schrödinger equation that the wave function $\psi_{(-)}$ obeys. The transformation (7) actually made use of the NKE branch $E_{(-)}$ and the resulted NKE Schrödinger equation for the $\psi_{(-)}$ applies to regions $E < V$. The kinetic energy in the NKE Schrödinger equation has always the form of (5). In the Appendix of this paper, how the transformations (6) and (7) lead to their resultants is presented in detail. We proposed experiments to verify the existence of NKE electrons.[1]

One of the author's basic viewpoint is that the NKE and PKE solutions ought to be treated on an equal footing. The RQMEs were of symmetry with respect to the PKE and NKE, e.g., Eq. (3). This symmetry should remain when low-momentum approximations of the RQMEs were taken. The transformations (6) and (7) are symmetric ones. The combination of the Schrödinger equation and NKE Schrödinger equation inherited the PKE-NKE symmetry of the RQMEs:[1] the Schrödinger equation and NKE Schrödinger equation were symmetric with respect to the PKE and NKE; the directions of the probability currents of the PKE and NKE solutions for free particles were opposite in direction; when potential changes a negative sign, the eigenvalues do either but the corresponding eigenfunctions remained unchanged, (see Appendix).

As soon as the status of the NKE was promoted to the same as that of the PKE, the fundamental equations in QM were inspected carefully.[5] The field of QM was extended to the domain of the NKE, and a series work needed to be done. In Appendix D of [1], 13 points were listed which were topics to be done in the following work. The points 2, 4, 6, 7 and 9 were dealt with in [2-5]. The present paper intends to deal with point 13. In [6], the Virial theorem was investigated comprehensively and the symmetry of the PKE and NKE was well exhibited in this theorem. In [7] the author resolved a few one-dimensional problems usually treated in QM textbooks, disclosing the discrepancies arising from taking into account the NKE Schrödinger equation.

The NKE branch puzzled people very much since it had never been met in reality. So, people have been reluctant to accept it. The NKE solutions were explained as antiparticles by Dirac as soon as he obtained them. This explanation has been retained ever since then. Such an explanation actually transferred the negative sign in energy expression (4) to be that in electric charge, and would inevitably raise contradictions, as argued in the Appendices in [1] and [5].

Our point of view is that the NKE solutions represent dark particles instead of antiparticles and that the NKE substances are dark matter (DM) that people have been looking for. The author regards NKE matter and DM as synonym. Dark particles are all what we have already known, and they are dark to us when they are of NKE. When a particle is of PKE, it can be detected by apparatus, and if it is of NKE, it cannot be.

Then a question emerges: why the NKE substances are dark. This question reminds us the reason why the matter is not dark, i.e., how we can observe a system, such as a remote celestial body.

It was analyzed[3] that a feature of a PKE system is that its energy spectrum has a lower limit but no upper limit, and the temperature of such a system must be positive. In a PKE system, the probabilities of particles occupying lower energy levels are larger than higher levels. The lowest energy level is the ground state, i.e., the stablest state. A particle in a lower energy state can absorb photons to transit to a higher energy one, which is called stimulated transition. This process cannot be spontaneous. Reversely, a particle in a higher level can emit photons to transit to a lower one, called spontaneous transition. Spontaneous transition occurs only when the occupying probability of the initial state is less than the final state. Photons, as quanta of the radiation field, have positive energies. The energy spectrum of the radiation has a lower limit but no upper limit, so that radiation is of positive temperature. Therefore, PKE systems absorb and emit positive energy, and both are of positive temperature. It is concluded that in our universe, PKE matter matches with energy. The emitted photons can be detected by us, because all of our devices are made of PKE particle. In this sense, the PKE systems are observable to us.

Now, we turn to NKE systems. A feature of a NKE system is that its energy spectrum has an upper limit but no lower limit, and the temperature of such a system is necessarily negative.[3] In a NKE system, the probabilities of particles occupying higher energy levels are larger than lower levels. The highest energy level is the ground state, i.e., the stablest state. These properties are just contrary to those of PKE systems. In a NKE system, particles certainly can transit between energy levels. Then how can the transitions be implemented? Intuitively, it seems that a particle in a higher level can emit energy to transit to a lower one, but this cannot be spontaneous, because the occupying probability of the initial state is greater than the final one. Spontaneous transition can occur only from a lower energy state to a higher one. However, if so, the NKE needs to absorb energies. Therefore, it is a severe problem how these transitions can happen. Besides, the NKE systems are of negative temperature, while radiation is of positive temperature. We say that NKE matter does not match energy. One of our tasks is to figure out how the NKE systems transit from a state to another.

Personally, the spontaneous transition from the lower states to the higher ones can no doubt occur in NKE systems. The poser is that the energy of the initial state minus that of the final state is negative. Then, logically, it must be that in the course of the spontaneous transition, the NKE system releases negative energy. Reversely, when a NKE system is in a higher level, it can absorb negative energy to transit to a lower level. The author's belief is that the negative energy is just dark energy (DE). Thus, we are going to put forth a new version of DE.

The DE was conceived in 1998.[8] By observation, the expansion of our universe has an acceleration. People conceived that some unobservable energy, called DE, was responsible for the accelerated cosmic expansion. There have been several versions of the DE.[9-14] It was thought that the DE was vacuum energy[15] and believed that the DE could couple with neutrinos.[16] The method of probing DE was suggested,[17] and a great amount of data concerning DE has been surveyed.[18] In spite of these efforts, the DE itself remains open. Among various scenarios, the most famous one might be the "chameleon field."[19] However, by experiments, "chameleon fields may have to be ruled out as sources of dark energy."[20]

The present work provides a new possible scenario of DE. From the analysis above, it is recognized that NKE system releases and absorbs negative energies to implement the transitions. Then there must be photons with negative energies, which are called negative photons. Since negative energy is DE, the negative photons are dark photons. Personally, realizing DM and DE depends on the conceptual breakthrough. If the NKE is a conceptual breakthrough for DM, the dark photon is the one for DE. The DM releases and absorbs DE. So, DM and DE match, just as that matter and energy match as mentioned above.

The task of the present work is to give the mathematical description and possible physical properties of the envisaged DE.

In Section II, we first briefly review the natural number photon states (NNPSs), and then extend them to negative integer photon states (NIPSs). In Section III, the formalism of statistical mechanics and thermodynamics of the dark photon field are presented. In Section IV, the vacuum zero-point energy (VZPE) in the universe is discussed. Section V is our conclusions. In Appendix, it is shown how particle number states are extended from the positive integer ones to negative integer ones.

## II. PHOTONS WITH MEGATIVE ENERGIES IN NEGATIVE INTEGER PHOTON STATES

We first carefully inspect the two energy branches in Eq. (3). In Eq. (2), each of the two terms in the square root can be zero. As the momentum is zero,

$$E_{(+)} = mc^2. \tag{8}$$

This is the famous mass-energy relationship derived by Einstein. This relationship has dual meanings. One is that a particle's mass $m$ is equivalent to a certain amount of static energy. The transformation (6) gets rid of this positive static energy from the RQMEs so as to achieve the Schrödinger equation. The other meaning is that a mass $m$ and a certain amount of energy can mutually converse, which has been verified by experiments, such as nuclear fission and fusion. One more example is that a pair of electron and positron can convert to a pair of photons and vice versa.

As the mass in (2) is zero,

$$E_{(+)} = cp. \tag{9}$$

This is the energy-momentum relationship of massless photons.

Just as Eq. (2), each term in the square root in (4) can be zero. As the momentum is zero,

$$E_{(-)} = -mc^2. \tag{10}$$

This is another mass-energy relationship. It has also dual meanings. One is that a particle's mass $m$ is equivalent to a certain amount of negative energy. The transformation (7) removes this negative static energy from the RQMEs so as to gain the NKE Schrödinger equation. The other meaning is that a mass $m$ and a certain amount of negative energy can mutually converse, the physical implications of which will be explored in our future work.

As the mass in (4) is zero,

$$E_{(-)} = -cp. \tag{11}$$

This is the energy-momentum relationship of massless photons with negative energies, i.e., dark photons.

There was another argument supporting Eq. (11). It was discussed[21] that relativity required that a zero-mass particle's four-vector $p = (\boldsymbol{p}, iE/c)$ should meet the energy-momentum relation

$$E^2 = c^2 p^2. \tag{12}$$

This formula contains both (9) and (11). It means that a photon's energy can be either positive or negative.

Now that the negative energy is composed of dark photons, it is desirable to give the representation of the negative photons. Before doing so, let us recall how to mathematically describe the energy, or radiation.

Free electromagnetic field can be quantized,[21-26] and the quanta are called photons. The energy of a photon is

$$E = \hbar \omega, \tag{13}$$

where the $\omega$ is the frequency of the electromagnetic field. Equation (13) is equal to (9) since there is wavelength-momentum de Broglie relation

$$\lambda = \frac{\hbar}{p}. \tag{14}$$

The radiation with frequency $\omega$ can be represented by photon number states (PNSs). A PNS is denoted by $|n\rangle$ where $n$ is any natural number, $n \geq 0$. Because of this reason, a PNS is called NNPS. In such a state, there can be $n$ photons, and $n = 0$ means there is no photon, the vacuum state. Such a system is also called photon field.

The actions of the photon's annihilation and creation operators on a NNPS is to raise and to lower the state, respectively,

$$a|n\rangle = \sqrt{n}|n\rangle, a^+|n\rangle = \sqrt{n+1}|n\rangle. \tag{15}$$

Especially,

$$a|0\rangle = 0. \tag{16}$$

Their commutator is

$$aa^+ - a^+a = 1. \tag{17}$$

The NNPSs are the eigenstates of the photo number operator (PNO),

$$\hat{n} = a^+a, \tag{18}$$

and

$$\hat{n}|n\rangle = n|n\rangle. \tag{19}$$

By use of (15), the state $|n\rangle$ can be expressed by

$$|n\rangle = \frac{1}{\sqrt{n!}} a^{+n} |0\rangle. \tag{20}$$

This demonstrates that $n$ quanta are created from the state with null photons. The NNPSs are orthonormal to each other,

$$\langle m|n\rangle = \delta_{mn} \tag{21}$$

and form a complete set,

$$\sum_{n=0}^{\infty} |n\rangle\langle n| = 1. \tag{22}$$

The Hamiltonian of the photon field with frequency $\omega$ in the photon number representation is expressed by

$$H_{(+)} = (a^+a + \frac{1}{2})\hbar\omega. \tag{23}$$

The energy spectrum has a lower limit $\hbar\omega/2$ and no upper limit. The ground state is $|0\rangle$ as revealed by Eq. (16). The above formulas are listed in TABLE I. We say that the NNPS constitute a positive photon system (PPS).

Now we extend the $n$ to be also negative integers, and such states are named as NIPSs. The NIPSs compose a negative photon system (NPS) which is called negative radiation (NR).

It is seen from the course of the quantization of electromagnetic field, each mode of the field is equivalent to a harmonic oscillator.[21-23] The Hamiltonian of the radiation field in the photon number representation, Eq. (23), has the same form of that of a harmonic oscillator in particle number representation.

In Appendix, it is shown how the positive integer states in the particle number representation of the harmonic oscillator system are extended to negative integer states. For photon number states, we can do the same thing.

Hence, the photon number states $|n\rangle$, where $n$ are negative integers, are called NIPS. For convenience, the NIPSs are also denoted by $|-n\rangle, n \geq 0$. The physical meaning of the $|-n\rangle, n \geq 0$ is that there are $n$ dark photons. When the frequency is $\omega$, each dark photon is of energy $-\hbar\omega$. Please note that the concept of dark photons here is different from that of "dark photon" in some literature;[27,28] the negative energy does not mean negative frequency,[29] since in the present work the frequency is positive but energy is negative.

We define a pair of operators $b$ and $b^\times$. They act on the NIPSs in the following manner:

$$b|-n\rangle = \sqrt{-n}\,|-n+1\rangle, \quad b^\times|-n\rangle = \sqrt{-n-1}\,|-n-1\rangle; \quad n \geq 0. \tag{24}$$

When $b$ acts on a NIPS, it makes the dark photon number decrease by 1. We say that $b$ annihilates a dark photon in the NIPS. Similarly, the $b^\times$ creates a dark photon in a NIPS. The dark photons compose dark radiation field, which are the synonym of negative photon system and negative radiation field.

It follows from (24) that

$$|-n\rangle = \frac{b^{\times n}}{i^n \sqrt{n!}}|0\rangle, \quad n \geq 0. \tag{25}$$

Creating $n$ dark photons in the vacuum state $|0\rangle$ produces state $|-n\rangle$. The NIPSs are orthonormal to each other,

$$\langle -m|-n\rangle = \delta_{mn}, \tag{26}$$

and form a complete set,

$$\sum_{n=0}^{\infty} |-n\rangle\langle -n| = 1. \tag{27}$$

The Hamiltonian of the dark photon field with frequency $\omega$ in the NIPS representation is expressed by

$$H_{(-)} = (b^\times b - \frac{1}{2})\hbar\omega. \tag{28}$$

There is no overlap between the two sets. This can be understood by the actions of the operators on the vacuum state:

$$b|0\rangle = 0, a|0\rangle = 0, b^\times|0\rangle = \sqrt{-1}|-1\rangle, a^+|0\rangle = |1\rangle. \tag{29}$$

The formulas for the NIPSs are listed in TABLE I.

**TABLE I. The formulas of the natural number photon states (NNPSs) and negative integer photon states (NIPSs).**

| | | NNPSs $|n\rangle, n \geq 0$ | NIPSs $|-n\rangle, n \geq 0$ |
|---|---|---|---|
| The ground state | | $|0\rangle$, $a|0\rangle = 0$ | $|0\rangle$, $b|0\rangle = 0$ |
| PNS | | $|n\rangle = \dfrac{a^{+n}}{\sqrt{n!}}|0\rangle$ | $|-n\rangle = \dfrac{b^{\times n}}{i^n\sqrt{n!}}|0\rangle$ |
| The orthogonality between the PNSs | | $\langle m|n\rangle = \delta_{mn}$ | $\langle -m|-n\rangle = \delta_{mn}$ |
| The completeness of the set | | $\sum_{n=0}^{\infty}|n\rangle\langle n| = 1$ | $\sum_{n=0}^{\infty}|-n\rangle\langle -n| = 1$ |
| Annihilating a photon or a dark photon | | $a|n\rangle = \sqrt{n}|n-1\rangle$ | $b|-n\rangle = \sqrt{-n}|-n+1\rangle$ |
| Creating a photon or a dark photon | | $a^+|n\rangle = \sqrt{n+1}|n+1\rangle$ | $b^\times|-n\rangle = \sqrt{-n-1}|-n-1\rangle$ |
| The eigenvalue of photo number operator | | $a^+a|n\rangle = n|n\rangle$ | $b^\times b|-n\rangle = -n|-n\rangle$ |
| Hamiltonian | | $H_{(+)} = (a^+a + \frac{1}{2})\hbar\omega$ | $H_{(-)} = (b^\times b + \frac{1}{2})\hbar\omega$ |
| The expectation of the $H$ in a PNS | | $(n+\frac{1}{2})\hbar\omega$ | $(-n-\frac{1}{2})\hbar\omega$ |
| Energy spectrum | Upper limit | No | $-\hbar\omega/2$ |
| | Lower limit | $\hbar\omega/2$ | No |

By the way, the introduction of the NIPSs helps us clarify an inequality that was mistaken as an identity. A pair of phase operators was defined as $e^{i\varphi} = (\hat{n}+1)^{-1/2}a$ and $e^{-i\varphi} = a^+(\hat{n}+1)^{-1/2}$. It was believed that $e^{i\varphi}e^{-i\varphi}$ was an identity operator but $e^{-i\varphi}e^{i\varphi}$ was not, i.e., $e^{i\varphi}e^{-i\varphi} = 1$ and $e^{-i\varphi}e^{i\varphi} \neq 1$, a pair of asymmetric relations.[30] This conclusion is questionable: since $(e^{i\varphi}e^{-i\varphi})^T = e^{-i\varphi}e^{i\varphi}$, the establishment of such a pair of asymmetric relations was unreasonable. Now that the NIPSs emerge, and the operators and are modified to be $e^{i\varphi}e^{-i\varphi} = -(\hat{n}+1)^{-1/2} bb^{\times}(\hat{n}+1)^{-1/2}$ for negative $n$. It is immediately seen that $(\hat{n}+1)^{-1/2} bb^{\times}(\hat{n}+1)^{-1/2} \neq 1$ as $n = -1$. We conclude that the correct relations are that $e^{i\varphi}e^{-i\varphi} \neq 1$ and $e^{-i\varphi}e^{i\varphi} \neq 1$.

## III. STATISTICAL MECHANICS AND THERMODYNAMICS OF THE NEGATIVE PHOTON SYSTEM

### A. Statistical mechanics

The NNPS system is black-body radiation (BBR), or simply radiation, whose theory can be seen in [31]. The NIPS system is temporarily called negative radiation (NR) or dark radiation. Now, we consider the statistical mechanics and thermodynamics of the NR.

The ground state of the NIPS system is $|0\rangle$ which has the highest energy $-\hbar\omega/2$. The energy spectrum of the NR is of an upper limit but no lower limit. It was argued[3] that such a system had necessarily negative temperature (NT). The concept of NT was proposed long before, and theoretical and experimental research have been carried out.[31-50] The systems to implement the NT in experiments have been spin systems.[32-38] The spin systems are composed of massive particles, and their energy spectra have both an upper limit and a lower limit. The discussion of the NT systems for those composed of massive particles was presented in [3]. In the present work, we investigate the massless dark photon systems.

We first briefly review the formulism of the BBR.[31] For a specific frequency $\omega$, the energy of the $n$th PNS is

$$\varepsilon_n = n\hbar\omega. \tag{30}$$

As a Bosonic system, the thermodynamic potential is

$$\Omega_{(+)}(\omega) = -\frac{1}{\beta}\ln\sum_{n=0}^{\infty}(e^{\beta(\mu-\hbar\omega)})^n = \frac{1}{\beta}\ln(1-e^{\beta(\mu-\hbar\omega)}), \tag{31}$$

where

$$\beta = \frac{1}{k_B T}. \tag{32}$$

$T$ is temperature. The formula for evaluating the mean number of particles is

$$\bar{n} = -\left.\frac{\partial \Omega(\omega)}{\partial \mu}\right|_{\mu=0}. \tag{33}$$

After the derivative, the chemical potential takes zero because the photon systems are so. From Eqs. (31) and (33) it is easily calculated that the mean number of a photon field is

$$\bar{n}_{(+)} = \frac{1}{e^{\beta\hbar\omega}-1}. \tag{34}$$

For the NIPS system, the energy of $n$th state with frequency $\omega$ is

$$\varepsilon_{-n} = -n\hbar\omega. \tag{35}$$

The thermodynamic potential is calculated by

$$\Omega_{(-)}(\omega) = -\frac{1}{\beta}\ln\sum_{n=0}^{-\infty}(e^{-\beta(\mu-\hbar\omega)})^n = \frac{1}{\beta}\ln(1-e^{\beta(\hbar\omega-\mu)}). \tag{36}$$

The definition of $\beta$ is still (32), but we keep in mind that temperature is negative.

$$\bar{n}_{(-)} = -\frac{1}{e^{-\beta\hbar\omega}-1}. \tag{37}$$

The $\bar{n}_{(-)}$ varies with temperature monotonically. As $-\beta\hbar\omega \gg 1$, $\bar{n}_{(-)} \approx 0$. That is to say, very much small $|T|$ means that the dark photons are rare. When $-\beta\hbar\omega \ll 1$, $\bar{n}_{(-)} \approx k_B T/\hbar\omega$. Large $|T|$ leads to large $|\bar{n}_{(-)}|$ and the latter is inversely proportional to frequency. The mean numbers of particles for the NNPS and NIPS systems are symmetric.

The above evaluation is based on the energy spectra (30) and (35). The actual Hamiltonian for a photon system is (23) which was obtained by quantization of the electromagnetic field.

The eigenenergies of the Hamiltonian (23) in PNS are

$$\varepsilon_n = (n+\frac{1}{2})\hbar\omega. \tag{38}$$

There is a zero-point energy. Usually, in textbooks of statistical mechanics, the zero-point energy was neglected.[31,51-53]

Now, we employ the energy levels (38) to calculate thermodynamic potential which is, in turn, expressed by partition function.

$$\Omega(\omega) = -\frac{1}{\beta}\ln Z(\omega). \tag{39}$$

For the photon system, the partition function is

$$Z_{(+)}(\omega) = \sum_{n=0}^{\infty} e^{\beta(n\mu-(n+1/2)\hbar\omega)} = \frac{e^{\beta\mu/2}}{2\sinh(\beta(\hbar\omega-\mu)/2)}. \tag{40}$$

Employing (39), (40), and (33), we obtain the particle number (34). Note that the zero-point energy is irrelative to chemical potential.

The averaged energy of frequency $\omega$ is

$$E_{(+)}(\omega) = \frac{1}{Z}\sum_{n=0}^{\infty} e^{-\beta\varepsilon_n}\varepsilon_n = -\frac{\partial}{\partial\beta}\ln Z_{(+)} = \frac{\hbar\omega}{e^{\beta\hbar\omega}-1} + \frac{1}{2}\hbar\omega. \tag{41}$$

The zero-point term $\hbar\omega/2$ is usually excluded[54] in discussing the thermodynamics of the photon system.

We turn to consider the dark photon system, the partition function of which is

$$Z_{(-)}(\omega) = \sum_{n=0}^{-\infty}(e^{\beta(n\mu-(-n-1/2)\hbar\omega)}) = \frac{e^{\beta\mu/2}}{2\sinh(-\beta(\hbar\omega-\mu)/2)}. \tag{42}$$

The calculated mean number of particles is the same as (37).

**B. Thermodynamics**

We are now at the stage to give the formalism of the thermodynamics of the NR. From the partition function (42), it is easy to calculate the averaged energy of frequency $\omega$ to be

$$E_{(-)}(\omega) = -\frac{\partial}{\partial\beta}\ln Z_{(-)} = -\frac{\hbar\omega}{e^{-\beta\hbar\omega}-1} - \frac{1}{2}\hbar\omega, \beta < 0. \tag{43}$$

This is just the opposite number[b] of (41). We denote

$$u_{(-)}(\omega) = -\frac{\hbar\omega}{e^{-\beta\hbar\omega}-1}. \tag{44}$$

The second term in (43) is left to be discussed in Section IV.

It is postulated that the density of mode per unit interval in $\omega$ in a unit volume is the same as that in the BBR,

$$\rho(\omega) = \frac{\omega^2}{\pi^2 c^3}. \tag{45}$$

Then the average energy density per unit interval in $\omega$ is that

$$\bar{U}_{(-)}(\omega) = \rho(\omega)u_{(-)}(\omega) = -\frac{\hbar\omega^3}{\pi^2 c^3}\frac{1}{e^{-\beta\hbar\omega}-1}. \tag{46}$$

It is just the opposite number of the energy density of BBR, and hence its mathematical properties are easily discussed by imitation of those of BBR. For examples, As $-\beta\hbar\omega \ll 1$, $\bar{U}_{(-)}(\omega) = \frac{\omega^2 k_B T}{\pi^2 c^3}$, which corresponds to "Rayleigh's law". This is independent of Plank's constant so that can be regarded as "classical limit". In other words, very large NT plays a role of "high temperature". On the other hand, as $-\beta\hbar\omega \gg 1$, $\bar{U}_{(-)}(\omega) = -\frac{\hbar\omega^3}{\pi^2 c^3} e^{\beta\hbar\omega}$. This is "Wien's law". The "Wien's displacement law" also applies.

---

[b] When two numbers have the same absolute value but are opposite in sign, either of them is called the opposite number of the other.

The integration of $\bar{U}_{(-)}(\omega)$ with respect to frequency is defined as the internal energy of the NR,

$$U_{(-)} = \int_0^\infty d\omega \bar{U}_{(-)}(\omega) = -\frac{\pi^2 k_B^4 T^4}{15 c^3 \hbar^3}. \tag{47}$$

Apparently, $U$ is just the opposite number of the internal energy of the BBR. Thus, it can be said that the Stefan-Boltzmann law still applies, i.e., the internal energy is proportional to the four powers of temperature.

From the relationship

$$U = F - TS = F - T\frac{\partial F}{\partial T} = -T^2 \frac{\partial}{\partial T}\frac{F}{T}, \tag{48}$$

the free energy can be calculated from the internal energy,

$$F = -T\int_0^T \frac{U}{T'^2} dT'. \tag{49}$$

Substitution of (47) into (49) results in

$$F_{(-)} = \frac{\pi^2 k_B^4 T^4}{45 c^3 \hbar^3}. \tag{50}$$

The free energy is a positive number. The relation $U = 3F$ still stands for the NR just as for the BBR.

The entropy is that

$$S_{(-)} = -\frac{\partial F_{(-)}}{\partial T} = -\frac{4\pi^2 k_B^4 T^3}{45 c^3 \hbar^3}. \tag{51}$$

The entropy is a positive number and grows with increasing absolute value of NT. It is verified that these thermodynamic quantities meet the relation $F = U - TS$.

The $U$, $F$ and $S$ evaluated above are in a unit volume. Multiplying a volume, denoted by $V$, to the free energy, we are able to calculate pressure.

$$P_{(-)} = -(\frac{\partial F_{(-)}}{\partial V})_T = -\frac{\pi^2 k_B^4 T^4}{45 c^3 \hbar^3}. \tag{52}$$

The relation $P = U/3$ remains valid for the NR. It is seen that the NR generates a negative pressure. The negative pressure is believed responsible for the acceleration of the university inflation.[55]

We compare in TABLE II the properties of the black-body radiation and dark black-body radiation. The remarkable features are that the pressure and temperature of the dark photon system are negative. In TABLE III of [3], the physical properties of the PKE and NKE systems composed of massive particles were listed.

**TABLE II. Comparisons of properties of black-body radiation and dark black-body radiation.**

|  | Black-body radiation | Dark black-body radiation |
|---|---|---|
| The energy of a photon with frequency $\omega$ | $\hbar\omega$ | $-\hbar\omega$ |
| The ground state of the PNSs | $\|0\rangle$ | $\|0\rangle$ |
| Temperature | $T>0$, $\beta=1/k_\mathrm{B}T>0$ | $T<0$, $\beta=1/k_\mathrm{B}T<0$ |
| Partition function | $\dfrac{1}{2\sinh(\beta\hbar\omega/2)}$ | $\dfrac{1}{2\sinh(-\beta\hbar\omega/2)}$ |
| Average energy of frequency $\omega$ | $\dfrac{\hbar\omega}{e^{\beta\hbar\omega}-1}+\dfrac{1}{2}\hbar\omega$ | $-\dfrac{\hbar\omega}{e^{-\beta\hbar\omega}-1}-\dfrac{1}{2}\hbar\omega$ |
| Zero-point energy of frequency $\omega$ | $\dfrac{1}{2}\hbar\omega$ | $-\dfrac{1}{2}\hbar\omega$ |
| Thermodynamic potential $\Omega$ | $\dfrac{1}{\beta}\ln[2\sinh(\beta\hbar\omega/2)]>0$ | $\dfrac{1}{\beta}\ln[2\sinh(-\beta\hbar\omega/2)]<0$ |
| Free energy $F$ | $-\dfrac{\pi^2 k_\mathrm{B}^4 T^4}{45 c^3 \hbar^3}<0$ | $\dfrac{\pi^2 k_\mathrm{B}^4 T^4}{45 c^3 \hbar^3}>0$ |
| Internal energy $U$ | $\dfrac{\pi^2 k_\mathrm{B}^4 T^4}{15 c^3 \hbar^3}>0$ | $-\dfrac{\pi^2 k_\mathrm{B}^4 T^4}{15 c^3 \hbar^3}<0$ |
| Entropy $S$ | $\dfrac{4\pi^2 k_\mathrm{B}^4 T^3}{45 c^3 \hbar^3}>0$ | $-\dfrac{4\pi^2 k_\mathrm{B}^4 T^3}{45 c^3 \hbar^3}>0$ |
| Pressure $P$ | $\dfrac{\pi^2 k_\mathrm{B}^4 T^4}{45 c^3 \hbar^3}>0$ | $-\dfrac{\pi^2 k_\mathrm{B}^4 T^4}{45 c^3 \hbar^3}<0$ |
| Specific heat $C_\mathrm{V}$ | $\dfrac{4\pi^2 k_\mathrm{B}^4 T^3}{15 c^3 \hbar^3}>0$ | $-\dfrac{4\pi^2 k_\mathrm{B}^4 T^3}{15 c^3 \hbar^3}>0$ |

## IV. DISCUSSION

### A. Vacuum zero-point energy

Here, we make some discussion about the zero-point energy. It was assumed that there is a vacuum zero-point energy (VZPE) in the universe. Because the VZPE is related to the cosmological constant in Einstein field equations of general relativity, it should be treated cautiously. Theoretically, the VZPE was estimated by

$$U_{VZPE} = \frac{\hbar c}{4\pi^2} \int_0^{k_{max}} dk k^2 \sqrt{k^2 + m^2/\hbar^2}. \quad (53)$$

The value estimated in such a way was about $10^{120}$ times of the value by astronomical observation[56], which caused the cosmological constant problem[57,58]. Although scenarios of the elimination of cosmological constant were suggested[59,60] and measurements were carried on to testify the theories,[61,62] this problem remains intractable.[63]

In the author's opinion, the VZPE should be divided into two parts. One is for massive particles and the other for massless photons. Equation (53) is actually for massive particles, as the mass $m$ appears in the right hand side. We have investigated the topic of the VZPE for massive particles.[64] Actually, there were contradictory presentations for the zero-point energies of free particles and harmonic oscillators for relativistic and low-momentum cases, (see TABLE I in [64]). We clarified that there was no zero-point energy for the harmonic oscillators.[64] The case of free particles, which involves the second quantization, will be dealt with in our future work. Here, we merely discuss the part of massless photons.

We have shown above that in the universe there can be photons with positive as well as negative energies. The former is the NNPS system and the latter the NIPS system. Their averaged energies of frequency $\omega$ were calculated, see Eqs. (41) and (43). It is seen that both the radiation and NR have zero-point energies which are just opposite numbers, so that they eliminate each other. The conclusion is that the total VZPE for massless photons in the universe is zero.

It is stressed that here we are talking about the total VZPE in the whole space. The NNPS and NIPS systems can have their own zero-point energies. The null VZPE in the whole universe is resulted from the offset of the two opposite zero-point energies. In a confined room, there may appear some effect of the zero-point energy after some constrain is applied. Furthermore, nowadays, people can detect the effect of zero-point energy of photons with positive energies. The famous example was the Casimir-Polder force.[65-68]

**B. The match of dark matter and dark energy**

Following the theory of dark photons presented above, we are now able to answer the question raised in Introduction that how a NKE system transits from one state to another. The system in a lower energy level state can spontaneously transit to a higher level one by releasing dark photons. Reversely, stimulated transition from a higher level to a lower one occurs when the system absorbs dark photons. Such transitions can occur not only because the total energy is conserved, but also both the NKE system and dark photons are of negative temperature. We say that DM and DE match.

A PKE system cannot release and absorb dark photons to implement transitions between energy levels. When spontaneous transition occurs, it will necessarily emit photons with positive energies, and only the photons can start stimulated transitions. There is no way for a PKE system to release or to absorb dark photons. That is to say, matter and DE do not match. Similarly, DM and energy do not match.

The substances visible to us, such as human bodies and the devices made by human being, are all composed of PKE particles, such that they are all PKE systems. They cannot absorb or release dark photons. Therefore, the dark photons are really dark to us. Furthermore, because the NKE systems can only release and absorb dark photons, they are dark to us. Thus, we give the reasons why the NKE systems and negative energy photons are dark to us.

The NKE particles may constitute complex or even macroscopic systems, and the laws of motion were developed before.[4] It is also possible that the dark systems can evolve into organisms, or even into life. If so, they can observe their environment. However, because they can merely absorb and release dark photons, they are unable to detect photons emitted from matters with PKE. Therefore, to them, all the PKE substances and energies are dark.

It is seen that the dark radiation produces negative pressure. It was mentioned in Introduction that people thought DE, although it was still unknown, produced negative pressure which was responsible for the accelerate inflation of our universe. It is disclosed in the present work that the negative radiation naturally produces negative pressure, which support our viewpoint that the photons with negative energies are DE. Furthermore, the NKE systems, i.e., DM, also produce negative pressure[3] and the kinetic mechanism was analyzed,[4] which seemed not conceived by others. Matter and energy produce positive pressure, and DM and DE produce negative pressure.

In the universe, there are symmetries with respect to PKE and NKE matters, positive and negative photon energies, the matter-energy match and DM-DE match.

## V. CONCLUSIONS

In this paper, the eigenstates of the photon are extended to the cases where the photon number can be negative. The physical meaning of negative photon number states is that a photon has a negative energy, called negative photon or dark photon. The dark photons compose negative radiation or dark radiation. The formulism of statistical mechanics and thermodynamics of the negative radiation is presented. The thermodynamic potential, internal energy, free energy, pressure are simply the opposite numbers of those of radiation, whereas the entropy and specific heat are positive.

The negative radiation is believed dark energy. Therefore, this work actually presents a theory of dark energy.

The dark energy is of negative pressure which is thought responsible for the accelerated inflation of the universe.

Dark energy is of negative temperature. This coincides with systems with negative kinetic energy which were believed dark matter. The energy spectrum of a believed dark matter has an upper limit but no lower one. Therefore, the dark matters can absorb and emit dark energy. This is just the opposite to the cases of known matter with positive kinetic energies and photons with positive energies. In this view, we say that matter and energy match and dark matter and dark energy match.

The universe is of symmetries of positive kinetic energy-negative kinetic energy, of matter-dark matter, of energy-dark energy, of matter-energy match and dark matter-dark energy match.

## ACKNOWLEDGMENTS
This work is supported by the National Natural Science Foundation of China under Grant No. 12234013, and the National Key Research and Development Program of China Nos. 2018YFB0704304 and 2016YFB0700102.

## APPENDIX A: THE EXTENSION OF PARTICLES NUMBER STATES FROM NATURAL NUMBERS TO THE WHOLE INTEGER NUMBERS.

We first recall the two low-momentum approximations of the Dirac equation.[1,5] This is to show that the Schrödinger equation and NKE Schrödinger equation have equivalent status. Then the particle number states corresponding to the Schrödinger equation are extended to the cases, corresponding to the NKE Schrödinger equation, where the number can be negative.

Dirac equation reads

$$i\hbar \frac{\partial}{\partial t}\Psi = H\Psi, \tag{A1}$$

where the Hamiltonian is

$$H = c\boldsymbol{\alpha} \cdot \boldsymbol{p} + mc^2\beta + V. \tag{A2}$$

Here $V$ means potential energy. The corresponding stationary equation is

$$(c\boldsymbol{\alpha} \cdot \boldsymbol{p} + mc^2\beta + V)\Psi = E\Psi. \tag{A3}$$

By the derivation procedure of the Dirac equation, we are aware of that the $\boldsymbol{\alpha}$ can also take a minus sign.[1] In (A3) we replace the $\boldsymbol{\alpha}$ with $-\boldsymbol{\alpha}$, and then change the sign of the whole equation. In this way, Eq. (A3) is recast to be

$$(c\boldsymbol{\alpha} \cdot \boldsymbol{p} - mc^2\beta - V)\Psi = -E\Psi. \tag{A4}$$

It is seen that Eqs. (A3) and (A4) are the same Dirac equation. The Hamiltonian in (A4),

$$H = c\boldsymbol{\alpha} \cdot \boldsymbol{p} - mc^2\beta - V, \tag{A5}$$

is the same one as (A2).

In the Dirac equation (A1) with the Hamiltonian (A2), we make transformation for the wave function,

$$\Psi = e^{-imc^2 t/\hbar}\psi_{(+)}, \tag{A6}$$

and take low-momentum approximation. Then, the wave function $\psi_{(+)}$ obeys the Schrödinger equation,

$$i\hbar \frac{\partial}{\partial t} \psi_{(+)} = H_{(+)} \psi_{(+)}, \tag{A7}$$

where

$$H_{(+)} = -\frac{\hbar^2}{2m} \nabla^2 + V. \tag{A8}$$

Equation (A7) with the Hamiltonian (A8) is well-known Schrödinger equation where the kinetic energy operator is $-\frac{\hbar^2}{2m}\nabla^2$ which is believed the PKE operator.

In Dirac equation (A1) with the Hamiltonian (A4), we make transformation

$$\Psi = e^{imc^2 t/\hbar} \psi_{(-)} \tag{A9}$$

and take low-momentum approximation. Then, the wave function $\psi_{(-)}$ obeys

$$i\hbar \frac{\partial}{\partial t} \psi_{(-)} = H_{(-)} \psi_{(-)}, \tag{A10}$$

where

$$H_{(-)} = \frac{\hbar^2}{2m} \nabla^2 - V. \tag{A11}$$

Here the kinetic energy operator is $\frac{\hbar^2}{2m}\nabla^2$, which is opposite to that in (A8), so that called NKE operator. Because of this reason, Eq. (A10) with the Hamiltonian (A11) is called NKE Schrödinger equation.[1,3-5]

It is seen that for a specific potential $V$, as long as there is a Schrödinger equation, there must be corresponding NKE Schrödinger equation with an opposite potential. This is because both equations are the low-momentum approximations of the Dirac equation.

The stationary equations of Eqs. (A7) and (A10) are respectively

$$(-\frac{\hbar^2}{2m} \nabla^2 + V) \psi_{(+)} = E \psi_{(+)} \tag{A12}$$

and

$$(-\frac{\hbar^2}{2m} \nabla^2 + V) \psi_{(-)} = -E \psi_{(-)}. \tag{A13}$$

The latter can be obtained by multiplying a minus sign to the former. It is easily seen that the eigenfunctions of Eqs. (A12) and (A13) are the same, and their eigenvalues are simply opposite to each other.

It is stressed that each of Eqs. (A12) and (A13) has specific physical meanings, so that they are physically independent of each other, in spite of that they mathematically differ in merely minus sign.

In our opinion, the NKE solutions of the Dirac equation are dark particles. Consequently, the solutions of the NKE Schrödinger equation are dark particles doing low-momentum motion. The experiments detecting NKE electrons were suggested.[1] The statistical mechanics and thermodynamics of NKE systems were studied,[3] and the equations of motion for macroscopic dark bodies were derived.[4]

The model of one-dimensional harmonic oscillator of the Schrödinger equation is a typical solvable problem that almost every QM textbook introduces. The eigenstates of this model can be written in the particle number representation, and the Hamiltonian can be written by particle creation and annihilation operators.[24,69-71] This procedure is also called second quantization. In the particle number states, the numbers are non-negative, the origination of which is that the kinetic energy in the Schrödinger equation is positive. In the following, we will first briefly review the process of how the second quantization states are produced. Then the states are extended to the cases where particle number can be negative integers.

The eigen equation of the model is

$$(-\frac{\hbar^2}{2m}\frac{\partial^2}{\partial x^2} + \frac{1}{2}m\omega^2 x^2)\varphi_n = E_n \varphi_n .\tag{A14}$$

Let

$$\alpha = \sqrt{\frac{m\omega}{\hbar}} .\tag{A15}$$

The solved eigenfunctions are

$$\varphi_n(x) = N_n e^{-\alpha^2 x^2/2} H_n(\alpha x), \quad N_n = (\frac{\alpha}{\sqrt{\pi}2^n n!})^{1/2} .\tag{A16}$$

The eigenvalues are

$$E_n = (n+\frac{1}{2})\hbar\omega, \quad n \geq 0 .\tag{A17}$$

One is able to define the annihilation and creation operators as follows.

$$a = \frac{1}{\sqrt{2}}(\alpha x + \frac{1}{\alpha}\frac{\partial}{\partial x}), a^+ = \frac{1}{\sqrt{2}}(\alpha x - \frac{1}{\alpha}\frac{\partial}{\partial x}) .\tag{A18}$$

He then easily calculates that

$$a\varphi_n(x) = \sqrt{n}\varphi_{n-1}(x), \quad a^+\varphi_n(x) = \sqrt{n+1}\varphi_{n+1}(x) .\tag{A19}$$

It follows that

$$aa^+ - a^+a = 1 \tag{A20}$$

and

$$a^+a\varphi_n = n\varphi_n. \tag{A21}$$

An eigenstate $\varphi_n$ can be expressed by $|n\rangle$ which means that there are $n$ particles occupying this state and is called particle number representation. Equations (A19) and (A21) are recast be as follows.

$$a|n\rangle = \sqrt{n}\,|n-1\rangle, \quad a^+|n\rangle = \sqrt{n+1}\,|n+1\rangle. \tag{A22}$$

$$a^+a|n\rangle = n|n\rangle. \tag{A23}$$

The $|n\rangle$ is the eigenstate of the operator $a^+a$ and the eigenvalue is $n$. Thus,

$$a^+a = n \tag{A24}$$

is the particle number operator. In this representation, the Hamiltonian in (A14) is converted to be the form

$$H_{(+)} = (a^+a + \frac{1}{2})\hbar\omega. \tag{A25}$$

The matrices of the annihilation and creation operators are put down as follows:

$$(a)_{m,n} = \sqrt{n}\,\delta_{m+1,n}, (a^+)_{m,n} = \sqrt{m}\,\delta_{m,n+1}; 0 \leq m, n < \infty. \tag{A26}$$

The matrix of the particle number operator is a diagonal one.

$$(a^+a)_{m,n} = n\delta_{m,n}. \tag{A27}$$

Please note that in each of the three matrices, the upper left corner is the (0,0) element, and rows down and columns to the right in order.

By comparison of (A12) and (A13), we immediately know that corresponding (A14), there must be a NKE equation as follows.

$$(\frac{\hbar^2}{2m}\frac{\partial^2}{\partial x^2} - \frac{1}{2}m\omega^2 x^2)\varphi_{-n} = E_{-n}\varphi_{-n}. \tag{A28}$$

The equation also reflects that a NKE particle can have a stationary motion when it is subject to a repulsive potential.[4,6] It is easily acquired that the eigenenergies of (A28) are simply opposite numbers of (A14).

$$E_{-n} = (-n - \frac{1}{2})\hbar\omega, n \geq 0. \tag{A29}$$

In Eq. (A29), the integers are $-n$. This is why the index in Eq. (28) are labeled by $-n$.

The eigenfunctions of (A28) are the same as (A16) in the sense that

$$\varphi_{-n}(x) = N_n e^{-\alpha^2 x^2/2} H_n(\alpha x), n \geq 0. \tag{A30}$$

For the states with negative indices, a new pair of operators is defined as

$$b = \sqrt{-1}a, b^\times = \sqrt{-1}a^+,\tag{A31}$$

where the $a$ and $a^+$ were defined by (A18). Obviously, we have

$$b\varphi_{-n} = \sqrt{-n}\varphi_{-n+1}, b^\times\varphi_{-n} = \sqrt{-n-1}\varphi_{-n-1}.\tag{A32}$$

It follows that

$$bb^\times - b^\times b = -1\tag{A33}$$

and

$$b^\times b = -n.\tag{A34}$$

Hence, the particle number states are extended to the cases where the number can be negative integers. Physically, the negative integer $-n$ means $n$ dark particles which are of NKE. Thus, the negative integer number states are dark particle states. The operators $b$ and $b^\times$ annihilates and creates one dark particle, respectively. The matrices of the annihilation and creation operators are put down as follows.

$$(b)_{m,n} = \sqrt{n}\delta_{m+1,n}, (b^\times)_{m,n} = \sqrt{m}\delta_{m,n+1}; -\infty < m, n \leq 0.\tag{A35}$$

The matrix of the particle number operator is a diagonal one.

$$(b^\times b)_{m,n} = n\delta_{m,n}.\tag{A36}$$

In each of the three matrices, the lower right corner is the (0,0) element, and rows up and columns to the left in order. The Hamiltonian in (A28) is expressed by

$$H_{(-)} = (b^\times b - \frac{1}{2})\hbar\omega.\tag{A37}$$

It is noted that here we have defined an operator $b^\times$ by (A31). The operators $b^\times$ and $b$ are not adjoint of each other, but they are pseudo-adjoint of each other. The definition of the concept of pseudo-adjoint and related mathematical theory will be given elsewhere. Here we merely emphasize that the operator $b^\times b$ is a self-adjont one, so that its eigenvalues are necessarily real, not restricted to be positive ones. In the present specific case, it suffices to mention that the matrices of $b^\times$ and $b$ happen to be transpose of each other and the eigenvalues of the matrix $b^\times b$ are real. The $b^\times$ and $b$ which act on the NIPSs undertake the same roles as $a^+$ and $a$ which act on the NNPSs.

**References**

[1] H.-Y. Wang, J. Phys. Commun. **4**, 125004 (2020).
https://doi.org/10.1088/2399-6528/abd0

[2] H.-Y. Wang, J. Phys. Commun. **4**, 125010 (2020).
https://doi.org/10.1088/2399-6528/abd340

[3] H.-Y. Wang, J. Phys. Commun. **5**, 055012 (2021).
https://doi.org/10.1088/2399-6528/abfe71

[4] H.-Y. Wang, J. Phys. Commun. **5**, 055018 (2021).
https://doi.org/10.1088/2399-6528/ac016b

[5] H.-Y. Wang, Physics Essays **35**(2), 152 (2022).
http://dx.doi.org/10.4006/0836-1398-35.2.152

[6] H.-Y. Wang, J. of North China Institute of Science and Technology **18**(4), 1 (2021) (in Chinese). http://19956/j.cnki.ncist.2021.04.001

[7] H.-Y. Wang, J. of North China Institute of Science and Technology **19**(1), 97 (2022) (in Chinese). http://10.19956/j.cnki.ncist.2022.01.016

[8] A. Cho, Science **336**, 1090 (2012). https://10.1126/science.336.6085.1090-b8

[9] G. Narain, T. Li, Phys. Rev. D **97**, 063006 (2018).
https://doi.org/10.1103/PhysRevD.97.063006

[10] Y. Cui, M. Pospelov, and J. Pradler, Phys. Rev. D **97**, 103004 (2018).
https://doi.org/10.1103/PhysRevD.97.103004

[11] N. Dalal, K. Abazajian, E. Jenkins, and A. V. Manohar, Phys. Rev. Lett. **87**, 141302 (2001). https://doi.org/10.1103/PhysRevLett.87.141302

[12] P. Creminelli and F. Vernizzi, Phys. Rev. Lett. **119**, 251302 (2017).
https://doi.org/10.1103/PhysRevLett.119.251302

[13] J. M. Ezquiaga and M. Zumalacarregui, Phys. Rev. Lett. **119**, 251304 (2017).
https://doi.org/10.1103/PhysRevLett.119.251304

[14] T. Josset and A. Perez, Phys. Rev. Lett. **118**, 021102 (2017).
https://doi.org/10.1103/PhysRevLett.118.021102

[15] H. Aoki, S. Iso, D. S. Lee, Y. Sekino, and C. P. Yeh, Phys. Rev. D **97**, 043517 (2018).
https://doi.org/10.1103/PhysRevD.97.043517

[16] N. Klop and S. Ando, Phys. Rev. D **97**, 083523 (2018).
https://doi.org/10.1103/PhysRevD.97.083523

[17] T. R. Schibli, Nat. Photonics **2**, 712 (2008).
https://doi.org/10.1038/nphoton.2008.240

[18] M. A. Troxel, et al., Phys. Rev. D **98**, 043528 (2018).
https://doi.org/10.1103/PhysRevD.98.043528

[19] J. Khoury and A. Weltman, Phys. Rev. D **69**, 044026 (2004).
https://doi.org/10.1103/PhysRevD.69.044026

[20] T. Jenke, Nat. Phys. **13**, 920 (2017). https://doi.org/10.1038/nphys4250

[21] V. B. Berestetskii, E. M. Lifshitz, and L. P. Pitaevskii, *Quantum Electrodynamics. Vol. 4 of Course of Theoretical Physics* (Pergmon Press, New York, 1982), Chap. 1.

[22] M. Q. Scully and M. S. Zubairy, *Quantum Optics* (Cambridge University Press, Cambridge, 1997), Chap. 1.

[23]D. F. Walls and G. J. Milburn, *Quantum Optics*, 2nd ed. (Springer-Verlag, Berlin, Heidelberg, 2008), Chap. 2.

[24]L. I. Schiff, *Quantum Mechanics* (McGraw-Hill Book Company, Inc., New York, 1949), Chaps 13 and 14.

[25]W. Greiner and J. Reinhardt, *Field Quantization* (Springer-Verlag, Berlin, 1996).

[26]A. Messiah, *Quantum Mechanics Vol. 2*, 2nd ed. (Dover Publications, Inc., Mineola, 1995), Chap. 21.

[27]J. P. Lees, et al., Phys. Rev. Lett. **119**, 131804 (2017).
https://doi.org/10.1103/PhysRevLett.119.131804

[28]R. Aaij, et al., Phys. Rev. Lett. **120**, 061801 (2018).
https://doi.org/10.1103/PhysRevLett.120.061801

[29]E. Rubino, et al., Phys. Rev. Lett. **108**, 253901 (2005).
https://doi.org/10.1103/PhysRevLett.108.253901

[30]G. Gerry and P. Knight, *Introductory Quantum Optics* (Cambridge University Press, Cambridge, 2005), pp. 25-34.

[31]L. D. Landau and E. M. Lifshitz, *Statistical Physics Part 1 Vol. 5 of Course of Theoretical Physics* (Pergmon Press, New York, 1980), pp. 158-166, 183-190, 221-224.

[32]R. V. Pound, Phys. Rev. **81**, 155 (1951). https://doi.org/10.1103/PhysRev.81.155

[33]N. F. Ramsey and R. V. Pound, Phys. Rev. **81**, 278 (1951).
https://doi.org/10.1103/PhysRev.81.278

[34]E. M. Purcell and R. V. Pound, Phys. Rev. **81**, 279 (1951).
https://doi.org/10.1103/PhysRev.81.279

[35]P. J. Hakonen, S. Yin, and O. V. Lounasmaa, Phys. Rev. Lett. **64**(22), 2707 (1990).
https://doi.org/10.1103/PhysRevLett.64.2707

[36]P. J. Hakonen, K. K. Nummila, R. T. Vuorinen, and O. V. Lounasmaa, Phys. Rev. Lett. **68**(3), 365 (1992). https://doi.org/10.1103/PhysRevLett.68.365

[37]P. J. Hakonen, R. T. Vuorinen, and J. E. Martikainen, Phys. Rev. Lett. **70**(18), 2818 (1993). https://doi.org/10.1103/PhysRevLett.70.2818

[38]A. S. Oja and O. V. Lounasmaa, Rev. Mod. Phys. **69**(1), 1 (1997).
https://doi.org/10.1103/RevModPhys.69.1

[39]N. F. Ramsey, Phys. Rev. **103**(1), 20 (1956).
https://doi.org/10.1103/PhysRev.103.20

[40]M. J. Kleun, Phys. Rev. **104**(3), 589 (1956).
https://doi.org/10.1103/PhysRev.104.589

[41]E. Abraham and O. Penrose, Phys. Rev. E **95** 012125 (2017).
https://doi.org/10.1103/PhysRevE.95.012125

[42]P. T. Landsberg, Phys. Rev. **115**(3), 518 (1959).
https://doi.org/10.1103/PhysRev.115.518

[43]B. D. Coleman and W. Noll, Phys. Rev. **115**(2), 262 (1959).
https://doi.org/10.1103/PhysRev.115.262

[44]C. E. Hecht, Phys. Rev. **119**(5), 1443 (1959).
https://doi.org/10.1103/PhysRev.119.1443

[45]A. P. Mosk, Phys. Rev. Lett. **95**, 040403 (2005).
https://doi.org/10.1103/PhysRevLett.95.040403

[46]A. Rapp, S. Mandt, and A. Rosch, Phys. Rev. Lett. **105**, 220405 (2010). https://doi.org/10.1103/PhysRevLett.105.220405

[47]Y. Yatsuyanagi, Y. Kiwamoto, H. Tomita, M. M. Sano, T. Yoshida, and T. Ebisuzaki, Phys. Rev. Lett. **94**, 054502 (2005). https://doi.org/10.1103/PhysRevLett.94.054502

[48]Y. Hama, W. J. Munro, and K. Nemoto, Phys. Rev. Lett. **120**, 060403 (2018). https://doi.org/10.1103/PhysRevLett.120.060403

[49]S. Hilbert, P. Hänggi, and J. Dunkel, Phys. Rev. E **90**, 062116 (2014). https://doi.org/10.1103/PhysRevE.90.062116

[50]H. Struchtrup, Phys. Rev. Lett. **120**, 250602 (2018). https://doi.org/10.1103/PhysRevLett.120.062116

[51]L. E. Reich, *A Modern Course in Statistical Physics* (University of Texas Press, Austin, 1980).

[52]K. S. Huang, *Statistical Mechanics* (John Wiley & Sons, New York, 1987).

[53]D. Chandler, *Introduction to Modern Statistical Mechanics* (Oxford University Press, New York, 1987).

[54]R. K. Pathria and P. D. Beale, *Statistical Mechanics*, 3rd ed. (Elsevier Ltd., Amsterdam, 2011), p. 200.

[55]J. Beringer, et al., (Particle Data Group) Phys. Rev. D **86**, 010001 (2012). https://doi.org/10.1103/PhysRevD.86.010001

[56]C. Zunckel and C. Clarkson, Phys. Rev. Lett. **101**, 181301 (2008). https://doi.org/10.1103/PhysRevLett.101.181301

[57]S. Weinberg, Rev. Mod. Phys. **61**, 1 (1989). https://doi.org/10.1103/RevModPhys.60.1

[58]M. Li, X. D. Li, S. Wang, and Y. Wang, Commun. Theor. Phys. **56**, 525 (2011).

[59]G. Gabadadze and S. Yu, Phys. Rev. D **98**, 024047 (2018). https://doi.org/10.1103/PhysRevD.98.024047

[60]J. P. Zibin, A. Moss, and D. Scott, Phys. Rev. Lett. **101**, 251303 (2008). https://doi.org/10.1103/PhysRevLett.101.251303

[61]G. Cronenberg, et al., Nat. Phys. **14**, 1022 (2018). https://doi.org/10.1038/s41567-018-0205-x

[62]M. Jaffe, et al., Phys. **13**, 938 (2017). https://doi.org/10.1038/nphys4189

[63]M. C. D. Marsh, Phys. Rev. Lett. **118**, 011302 (2017). https://doi.org/10.1103/PhysRevLett.118.011302

[64]H.-Y. Wang, Physics Essays, **35**(3), 270 (2022). http://dx.doi.org/10.4006/0836-1398-35.3.270

[65]H. B. G. Casimir and D. Polder, Phys. Rev. **73**(4), 360 (1948). https://doi.org/10.1103/PhysRev.73.360

[66]M. J. Sparnaay, Physica **24**(6-10), 751 (1958).

[67]L. Spruch and E. J. Kelsey, Phys. Rev. A **18**(3), 845 (1978). https://doi.org/10.1103/PhysRevA.18.845

[68]C. I. Sukenik, M. G. Boshier, D. Cho, V. Sandoghdar, and E. A. Hinds, Phys. Rev. Lett. **70**(5), 560 (1993). https://doi.org/10.1103/PhysRevLett.70.560

[69]A. Messiah, *Quantum Mechanics Vol. 1*, 2nd ed. (Dover Publications, Inc., Mineola, 1995), Chap. 12.

[70] C. Cohen-Tannoudji, B. Diu, and F. Laloë, *Quantum Mechanics* (I Hermann, éditeurs des scences et arts, 293 rue Lecourbe, 75015 Paris, 1973), Chap. 5.

[71] F. Scheck, *Quantum Physics*, 2nd ed. (Springer-Verlag: Berlin, Heidelberg, 2013). (eBook) http://10.1007/978-3-642-34563-0